\documentclass{jpsj3} 

\title{Occurrence of Fermi Pockets without Pseudogap Hypothesis 
\\ 
and Clarification of the Energy Distribution Curves of Angle-Resolved Photoemission Spectroscopy in Underdoped Cuprate Superconductors } 

\author{\name{Hiroshi \surname{Kamimura}\thanks{E-mail address: 
kamimura@rs.kagu.tus.ac.jp}} \name{Kenji \surname{Sasaoka}$^{1}$}, and \name{Hideki 
\surname{Ushio}$^{2}$} 
} 
\inst{\address{Research Institute for Science and Technology, Tokyo University of 
Science, 1-3 Kagurazaka, Shinjuku-ku, Tokyo 162-8601, Japan 
\\ 
$^{1}$The University of Tokyo, Department of Materials Engineering, 7-3-1 Hongo, Bunkyo-ku, Tokyo 113-8656, Japan} 
\\ 
$^{2}$Tokyo National College of Technology, 1220-2 Kunugida-chou, Hachioji 193-0997, 
Japan} 
\abst{Central issues in the electronic structure of underdoped cuprate superconductors 
are to clarify the shape of the Fermi surfaces and the origin of the pseudogap. 
On the basis of the model proposed by Kamimura and Suwa, which bears important features 
originating from the interplay of Jahn-Teller physics and Mott physics, the feature of Fermi 
surfaces in underdoped cuprates is the presence of Fermi pockets constructed from doped holes under 
the coexistence of a metallic state and a local antiferromagnetic order. Below $T_ 
{\rm c}$, the holes on Fermi pockets form Cooper pairs with d-wave symmetry in the nodal 
region. In the antinodal region, there are no Fermi surfaces. In this study we 
calculate the energy distribution curves (EDCs) of angle-resolved photoemission spectroscopy (ARPES) below $T_{\rm c}$. 
It is shown that the feature of ARPES profiles of underdoped cuprates consists of a coherent peak 
in the nodal region and real transitions of photoexcited electrons from occupied 
states below the Fermi level to a free-electron state above the vacuum level in the 
antinodal region, where the latter transitions form a broad hump. From this feature, the 
origin of the two distinct gaps observed by ARPES is elucidated without introducing the concept of the pseudogap. Finally, a remark is made on the phase diagram of underdoped cuprates.} 

\kword{underdoped cuprate superconductors, coexistence of 
AF order and a metallic state, Fermi pockets, d-wave superconductivity, theory 
of ARPES EDCs, Origin of 
broad hump in ARPES, phase diagram} 

\begin{document} 
\maketitle 

\section{Introduction} 
Undoped copper oxide (La$_2$CuO$_4$) is an antiferromagnetic Mott insulator, in which an 
electron correlation plays an important role \cite{Anderson}. Thus, we may say that 
undoped cuprates are governed by Mott physics. In 1986, Bednorz and M\"{u}ller discovered 
high-temperature superconductivity in copper oxides by doping hole carriers into La$_2 
$CuO$_4$ \cite{Bednorz}. Their motivation was the consideration that higher $T_{\rm c}$ could be achieved 
for copper oxide materials by combining Jahn-Teller (JT) active Cu ions with the 
structural complexity of layer-type perovskite oxides. To investigate the 
mechanism of high-temperature superconductivity, it is assumed in most models that doped 
holes itinerate through orbitals extending over a CuO$_{2}$ plane in systems 
consisting of CuO$_{6}$ octahedrons elongated by the JT effect. These models are 
called the ``single-component theory'', because the orbitals of hole carriers extend only 
over a CuO$_{2}$ plane. 

\begin{figure}[h] 
\begin{center} 
\includegraphics[width=9cm]{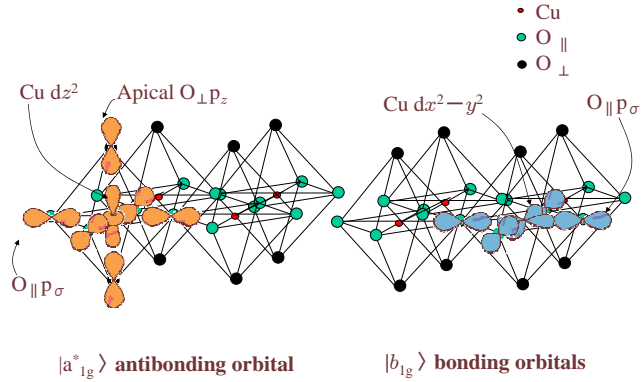} 
\end{center} 
\caption{\label{fig1} 
(Color online) Spatial extension of a$^*_{\rm 1g} $ antibonding orbital $|{\rm a}^*_{\rm 1g}\rangle $ 
and   b$_{\rm 1g} $ bonding orbital $|{\rm b}_{\rm 1g}\rangle $. {\textbf a,} a$^*_{\rm 
1g} $ antibonding orbital $|{\rm a}^*_{\rm 1g}\rangle $. {\textbf b,} b$_{\rm 1g} $ 
bonding orbital $|{\rm b}_{\rm 1g}\rangle $.} 
\end{figure}

In 1989, Kamimura and coworkers showed by first-principles calculations that the 
apical oxygen in CuO$_{6}$ octahedrons tends to approach Cu$^{2+}$ ions when 
Sr$^{2+}$ ions are substituted for La$^{3+}$ ions in La$_2$CuO$_4$ in order to gain the 
attractive electrostatic energy in ionic crystals such as cuprates \cite{Shima,Oshiyama}. 
As a result, CuO$_{6}$ elongated by the JT effect shrinks with hole doping. This 
deformation against the JT distortion is called the``anti-Jahn-Teller effect'' \cite 
{Copper_Oxide}. By this effect, the energy separation between the two kinds of orbital 
states, which have been split originally by the JT effect, becomes smaller with hole carrier doping. These two states are the a$_{\rm 1g}$ antibonding orbital state $|{\rm a}_{\rm 
1g}^*\rangle$ and ${\rm b}_{\rm 1g}$ bonding orbital state $|{\rm b}_{\rm 1g}\rangle 
$. The 
a$^*_{\rm 1g}$ antibonding orbital state is constructed by a Cu d$z^2$ orbital and the six 
surrounding oxygen p orbitals including apical O p$_{z}$-orbitals; the $|{\rm b}_{\rm 
1g}\rangle$ orbital is constructed by four in-plane O p$_{\sigma}$ orbitals with a small Cu d$_ 
{x^2-y^2}$ component parallel to the CuO$_2$ plane. 
The spatial extensions of the $|{\rm a}_{\rm 1g}^*\rangle$ and $|{\rm b}_{\rm 1g}\rangle 
$ orbitals, which are perpendicular and parallel to the CuO$_{2}$ plane, respectively, are 
shown in Fig.~\ref{fig1}. 

By taking account of the anti-Jahn-Teller effect, Kamimura and Suwa reported that one must 
consider these two kinds of orbital states equally in forming the metallic state of 
cuprates; they constructed a metallic state coexisting with the local 
antiferromagnetic (AF) order \cite{KS}. This model is called the ``Kamimura-Suwa (K-S) 
model'' \cite{Kamimura_Entropy}. Since the anti-Jahn-Teller effect is a central issue of 
Jahn-Teller physics, we may say that the K-S model bears important features originating from the interplay of Jahn-Teller physics and Mott physics. Since these two kinds of 
orbitals extend not only over the CuO$_{2}$ plane but also along the direction 
perpendicular to it,  the K-S model represents a prototype of a ``two-component theory'', 
in contrast to the single-component theory. 

On the basis of the K-S model, Kamimura and Ushio have calculated Fermi surfaces in underdoped 
La$_{2-x}$Sr$_x$CuO$_4$ (LSCO) \cite{KU, UK}, and have shown that the coexistence of a 
metallic state and a local AF order results in the Fermi pockets 
constructed from doped holes in the nodal region.  The appearance of Fermi pockets and 
small Fermi surfaces in cuprates has recently been reported by various experimental 
groups \cite{Meng, Nicolas_Doiron-Leyrud, Bangura, Yoshida1, Yoshida2, Charkravarty_Kee}. 

In this study, on the basis of the K-S model, we calculate the energy distribution curves (EDCs) of the angle-resolved photoemission spectroscopy 
(ARPES) profiles of cuprates below $T_{\rm c}$, and we show that the feature 
of the calculated ARPES profiles consists of a coherent peak 
due to the superconducting density of states in the nodal region and the real transitions 
of 
electrons from the occupied states below the Fermi level to a free-electron 
state above the vacuum level in the antinodal region. In particular, we show that the 
latter transitions form a broad hump in ARPES EDCs in underdoped cuprates. 

Concerning the ARPES experiments in underdoped cuprates, Tanaka and coworkers 
reported very interesting gap features in their observation of ARPES spectra. Their result 
exhibits a coherent peak in the nodal region and a broad hump 
in the antinodal region in underdoped Bi2212 samples below $T_{\rm c} $\cite{Tanaka}. 
From the quantitative agreement between the theory and the experiment, we conclude 
that the observed broad hump corresponds to the photoelectron excitations 
from the occupied states below the Fermi level to the free-electron state above the 
vacuum level. In this context, it is concluded that the introduction of the 
phenomenological idea of the pseudogap is not necessary. 
Finally, in connection with the finite size of the spin-correlation length in a 
metallic state, we discuss the finite size effect of a metallic state on the spin-electronic structures of underdoped cuprates, and a new explanation for the phase diagram 
for underdoped cuprates is proposed. 

 The organization of the present paper is as follows: At the 
beginning of \S 2 we first summarize the essential features of the K-S 
model, which bears important features originating from the interplay of JT 
physics and Mott physics. In \S 3, on the basis of the many-body effects including 
energy bands obtained from the K-S model, we predict the key features of ARPES 
EDCs and clarify the origin of the two-gap scenario proposed from the experimental 
results of Tanaka {\it et al} \cite{Tanaka}. In \S 4, we discuss the finite size 
effects on the Fermi surfaces in cuprates. In connection with the finite size effects, we 
discuss the possibility of the spatially inhomogeneous distribution of Fermi pocket states 
and large Fermi surface states.  Taking account of the finite size effect, we propose a new 
interpretation for the phase diagram of underdoped cuprates in \S 5. We devote \S 6 to 
the conclusion and concluding remarks. 

\section{On the K-S Model}
In this section, we summarize the main features of the K-S model \cite{KS}, emphasizing 
its important roles in underdoped cuprates due to the interplay of JT physics and Mott physics. 

\subsection{Key features of the K-S model}
\begin{figure} 
\begin{center} 
\includegraphics[width=9cm]{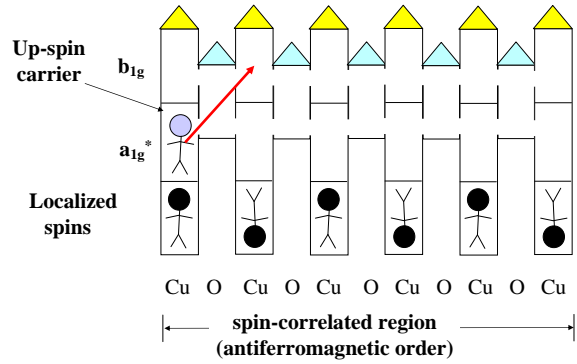}
\end{center} 
\caption{\label{fig2} 
(Color online) Explanation of K-S model using picture of two-story house model.} 
\end{figure}

The key features of the K-S model are explained in a heuristic way using the picture of a 
two-story house model shown in Fig.~\ref{fig2}. In this figure, the first story of a Cu 
house is occupied by Cu localized spins, which form the AF order in the spin-correlated region by the superexchange interaction. The second story in the Cu house 
consists of two floors due to the anti-JT effect; the lower a$^*_{\rm 1g}$ floor and the upper b$_{\rm 1g}$ floor. The second stories of neighboring Cu houses are connected by oxygen 
rooms, reflecting the hybridization of Cu d and O p orbitals. In the second story, a hole  
carrier with an up-spin enters the a$^*_{\rm 1g}$ floor of the left-hand Cu house owing 
to Hund's coupling with a Cu localized up-spin in the first story (Hund's coupling triplet) 
\cite{KE, EK}, as shown in the leftmost column of the figure. By the transfer 
interaction marked by a long arrow in the figure, the hole is transferred into the b$_ 
{\rm 1g}$ floor in the neighboring Cu house (second from the left) through the oxygen 
rooms, where a hole with up-spin forms a spin-singlet state with a localized down-spin 
in the second Cu house from the left (Zhang-Rice singlet) \cite{Zhang_Rice}. The key 
feature of the K-S model is that the hole carriers in the underdoped regime of LSCO form 
a metallic state by taking the Hund coupling triplet and the Zhang-Rice singlet 
alternately in the presence of a local AF order without destroying the AF order, as 
shown in the figure.  From Fig.~\ref{fig2}, one may understand that the characteristic 
feature of the K-S model is the coexistence of the AF order and a normal, metallic (or a 
superconducting) state in the underdoped regime. This feature of the K-S model (two-component theory) is different from that of the single-component theory. 

As seen in Fig.~\ref{fig2}, the wave functions of a hole carrier with up and down-spins 
have the following phase relation: 
   
\begin{eqnarray} 
\Psi _{\Vec{k}\downarrow} (\Vec{r} ) = \exp (i\Vec{k}\cdot \Vec{a} ) \Psi _{\Vec{k} 
\uparrow }(\Vec{r} ).                \label{eq:1} 
\end{eqnarray} 
Kamimura {\it et al.} have shown that this unique phase relation leads to the d-wave 
superconductivity \cite{Kamimura_d-Wave, Kamimura_Super}. 

\subsection{Effective Hamiltonian for the K-S model}
The following effective Hamiltonian is introduced to describe the K-S model 
following Kamimura and Suwa \cite{KS} (see also ref.~7). 
It consists of four parts: the one-electron Hamiltonian $H_{\rm sing}$ 
for the a$_{\rm 1g}^{\ast}$ and b$_{\rm 1g}$ orbital states, 
the transfer interaction between neighboring CuO$_6$ octahedrons $H_{\rm tr}$, 
the superexchange interaction between the Cu $d_{x^2-y^2}$ localized spins $H_{\rm AF}$, 
and the exchange interactions between the spins of dopant holes and $d_{x^2-y^2}$ 
localized holes 
within the same CuO$_6$ octahedron $H_{\rm ex}$. 
Thus, we have 
\begin{eqnarray} 
H &=& H_{\rm sing} + H_{\rm tr} + H_{\rm AF} + H_{\rm  ex}  \nonumber \\ 
 &=& \sum_{i,m,\sigma} \varepsilon_m C_{im\sigma}^\dagger C_{im\sigma} 
\nonumber \\ 
   & & {}+ \sum_{\langle i,j\rangle,m,n,\sigma} t_{mn} 
             \left(C_{im\sigma}^\dagger C_{jn\sigma} + {\rm h.c.} \right) \nonumber \\ 
 & & {} + J \sum_{\langle i,j\rangle} {\Vec{S}}_i \cdot {\Vec{S}}_j 
        + \sum_{i,m} K_m\, {\Vec{s}}_{i,m} \cdot {\Vec{S}}_i \ \ , \label{eq:2} 
\end{eqnarray} 
where $\varepsilon_m$ ($m=$ a$_{\rm 1g}^{\ast}$ or b$_{\rm 1g}$) 
represents the one-electron energy of 
the a$_{\rm 1g}^{\ast}$ and b$_{\rm 1g}$ orbital states, 
$C_{im\sigma}^\dagger$ and $C_{im\sigma}$ are the creation and annihilation operators of 
a dopant hole 
with spin $\sigma$ in the $i$th CuO$_6$ octahedron, respectively, 
$t_{mn}$ is the transfer integral of a dopant hole between the $m$-type and $n$-type orbitals 
of neighboring CuO$_6$ octahedrons, $J$ is the superexchange interaction 
between spins ${\Vec{S}}_i$ and ${\Vec{S}}_j$ of $d_{x^2-y^2}$ localized holes 
in the b$_{\rm 1g}^{\ast}$ orbital in the nearest-neighbor Cu sites $i$ and $j$ 
($J > 0$ for AF interaction), 
and $K_m$ is the exchange integral for the exchange interactions 
between the spin of a dopant hole, ${\Vec{s}}_{im}$, and the $d_{x^2-y^2}$ localized spin 
${\Vec{S}}_i$ 
in the $i$th CuO$_6$ octahedron. There are two exchange constants, i.e. $K_{\rm a_{\rm 1g}^ 
{\ast}}$  and 
$K_{\rm b_{\rm 1g}}$, for the Hund coupling triplet and the Zhang-Rice singlet, 
respectively, where 
$K_{\rm a_{\rm 1g}^{\ast}} < 0$ and $K_{\rm b_{\rm 1g}} > 0$. The appearance of the two 
kinds of 
exchange interactions in the fourth term is due to the interplay of Mott physics and 
JT physics. This is the key feature of the K-S model. 

The electron-electron interactions between doped hole carriers are very weak for two 
reasons: One is the low concentration of hole carriers in the underdoped regime and the 
other is the wave functions of hole carriers with up and down spins in a CuO$_6$ 
octahedron occupying a$_{\rm 1g}^{\ast}$ and b$_{\rm 1g}$ orbitals, respectively, as seen in 
eq.~(\ref{eq:1}) and Fig.~\ref{fig1}. For these reasons, we have neglected the 
electron-electron interactions between doped holes in the effective Hamiltonian~eq.~(2). 

By replacing the localized spins ${\Vec{S}}_i$'s in $H_{\rm ex}$ with their average 
$\langle {\Vec{S}}\rangle$ in the mean-field sense, we can calculate the change in the 
total energy upon moving a hole from an a$^*_{\rm 1g}$ orbital state in Hund's 
coupling spin triplet at Cu site $i$ to an empty b$_{\rm 1g}$ orbital state in the Zhang-Rice spin singlet at the neighboring Cu site $j$. In the first step, the hole moves from 
Cu site $i$ to infinity. The change in the total energy in the mean field approximation 
is equal to $\varepsilon_{{\rm a}_{\rm 1g}^{\ast}}+ \frac{1}{4}K_{{\rm a}_{\rm 1g}^ 
{\ast}}$. In the second step, the hole moves from infinity to an empty b$_{\rm 1g}$ 
orbital state at Cu site $j$ to form the Zhang-Rice singlet. The change in the total 
energy in the second step is equal to $\varepsilon_{\rm b_{\rm 1g}} - \frac{3}{4}K_{\rm 
b_{\rm 1g}}$. As a result, the change in the total energy by the transfer of the hole from 
the occupied ${\rm a_{\rm 1g}^{\ast}}$ orbital state at Cu site $i$ to the empty ${\rm b_ 
{\rm 1g}}$ orbital state at Cu site $j$ is 
\begin{eqnarray} 
 \varepsilon_{\rm a_{\rm 1g}^{\ast}}^{\rm ~eff} -  \varepsilon_{\rm b_{\rm 1g}}^{\rm 
~eff} = \varepsilon_{\rm a_{\rm 1g}^{\ast}} + \frac{1}{4}K_{\rm a_{\rm 1g}^{\ast}} - 
\varepsilon_{\rm b_{\rm 1g}} + \frac{3}{4}K_{\rm b_{\rm 1g}}\label{eq:D}.         
\end{eqnarray} 

Here, $\varepsilon_{\rm a_{\rm 1g}^{\ast}}^{\rm ~eff}$ and $\varepsilon_{\rm b_{\rm 1g}}^ 
{\rm ~eff}$ represent the effective one-electron energies of the a$_{\rm 1g}^{\ast}$ and b$_ 
{\rm 1g}$ orbital states including the exchange interaction term $H_{\rm  ex}$, 
respectively. Thus, the energy difference $(\varepsilon_{\rm a_{\rm 1g}^{\ast}}^{\rm ~eff} 
- \varepsilon_{\rm b_{\rm 1g}}^{\rm ~eff})$ corresponds to 
the energy difference between the ${\rm a_{\rm 1g}^{\ast}}$ and ${\rm b_{\rm 1g}}$ floors in 
the second story in Fig.~\ref{fig2}. 

\begin{figure}[h] 
\begin{center} 
\includegraphics[width=9cm]{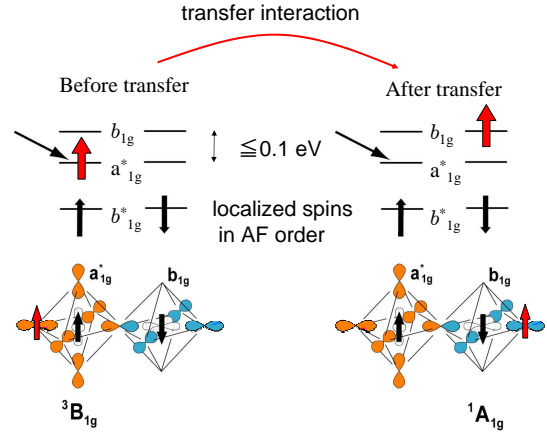} 
\end{center} 
\caption{\label{fig3} 
(Color online) Simple explanation for coexistence of metallic state and local AF order in 
two-component theory (a$^*_{\rm 1g}$ and b$_{\rm 1g}$ orbital states), by taking 
neighboring Cu sites $i$ to $j$ as an example, where localized spins are in 
antibonding b$^{\ast}_{\rm 1g}$ orbitals. Schematic pictures of Hund's coupling triplet $^ 
3$B$_{\rm 1g}$ and Zhang-Rice singlet $^1$A$_{\rm 1g}$ are also shown at the bottom. 
The energy in this figure is taken as the hole energy.} 
\end{figure} 

Now, let us estimate the energy difference $\varepsilon_{\rm a_{\rm 1g}^{\ast}}^{\rm 
~eff}$ -  $\varepsilon_{\rm b_{\rm 1g}}^{\rm ~eff}$ using the values of the 
parameters in the effective Hamiltonian~eq.~(\ref{eq:2}). The values of the parameters in 
Hamiltonian~eq.~(\ref{eq:2}) have been determined in the case of LSCO in ref.~6 (see 
also ref.~7). They are $J=0.1$, 
$K_{\rm a_{\rm 1g}^{\ast}}=-2.0$, $K_{\rm b_{\rm 1g}}=4.0$, 
$t_{\rm a_{\rm 1g}^{\ast} a_{\rm 1g}^{\ast}}=0.2$, 
$t_{\rm b_{\rm 1g} b_{\rm 1g}}=0.4$, 
$t_{\rm a_{\rm 1g}^{\ast} b_{\rm 1g}} 
= \sqrt{t_{\rm a_{\rm 1g}^{\ast} a_{\rm 1g}^{\ast}} t_{\rm b_{\rm 1g} b_{\rm 1g}}} \sim 
0.28$, 
$\varepsilon_{\rm a_{\rm 1g}^{\ast}}=0$, and $\varepsilon_{\rm b_{\rm 1g}}=2.6$ in units of 
eV, 
where $K_{\rm a_{\rm 1g}^{\ast}}$ and $K_{\rm b_{\rm 1g}}$ are taken from 
first-principles cluster calculations 
for a CuO$_{6}$ octahedron in LSCO \cite{KE, EK}, and the $t_{mn}$ are 
obtained by band structure calculation \cite{Shima, Oshiyama}. 
The difference in one-electron energy 
between the a$_{\rm 1g}^{\ast}$ and b$_{\rm 1g}$ orbital states in a CuO$_{6}$ octahedron for 
a certain $x$ has been determined so as to reproduce 
the difference in the lowest state energy between 
Hund's coupling spin-triplet state and the Zhang-Rice spin-singlet state for the same 
$x$ in LSCO calculated by Multi-Configuration Self-Consistent Field (MCSCF) 
cluster calculations which include the anti-JT effect \cite{KS}. 

Thus, the calculated $\varepsilon_{\rm a_{\rm 1g}^{\ast}}^{\rm ~eff}$ -   
$\varepsilon_{\rm b_{\rm 1g}}^{\rm ~eff}$ is 0.1 eV in the case of the optimum doping 
($x = 0.15$). Then, by introducing the transfer interaction of $t_{\rm a_{\rm 1g}^{\ast} 
b_{\rm 1g}} = 0.28$~eV, a coherent metallic state in the normal phase is obtained in the presence of the local AF order for the underdoped regime. This situation is 
schematically shown in Fig.~\ref{fig3}. 

\subsection{Features of the many-body effect including energy bands and Fermi surfaces of 
underdoped LSCO coexisting with the AF order}
In the previous subsection, we have shown that the effective Hamiltonian~eq.~(\ref{eq:2}) for 
the K-S model can lead to a unique metallic state in the normal phase, which results in 
the coexistence of a superconducting state and an AF order below $T_{\rm c}$. In 1994, 
Kamimura and Ushio calculated the energy bands and Fermi surfaces of underdoped LSCO in 
the normal phase on the basis of the effective Hamiltonian~eq.~(\ref{eq:2}), by treating the fourth 
term $H_{\rm ex}$ in the effective Hamiltonian~eq.~(\ref{eq:2}) by the mean-field 
approximation, that is, by replacing the localized spins ${\Vec{S}}_i$'s with their average 
$\langle {\Vec{S}}\rangle$ \cite{KU, UK}. Thus, the effect of the localized spin 
system was dealt with as an effective magnetic field acting on hole carriers. As a 
result, Kamimura and Ushio separated the localized hole-spin system in the AF order and the hole 
carrier system from each other, and calculated the ``one-electron type'' energy band for 
a carrier system assuming a periodic AF order. Here, ``one-electron type''  means the 
inclusion of many-body effects in the energy bands. That is, the effect of the exchange interactions 
between carriers and localized spins is included in the sense of the mean field approximation. 

In Fig. 4, the calculated many-body effect including energy band structure for 
up-spin (or down-spin) doped holes in LSCO is shown for various values of the wave vector 
$\Vec{k}$ and symmetry points in the AF Brillouin zone, where the AF 
Brillouin zone is adopted because of the coexistence of a metallic state and the AF 
order; it is shown on the left side of the figure. Here, note that the 
energy in this figure is taken as the electron energy but not as the hole energy. Furthermore, the 
Hubbard bands for localized b$^{\ast}_{\rm 1g}$  holes, which contribute to the local AF 
order, are separated from this figure and do not appear in this figure.

\begin{figure} 
\begin{center} 
\includegraphics[width=9cm]{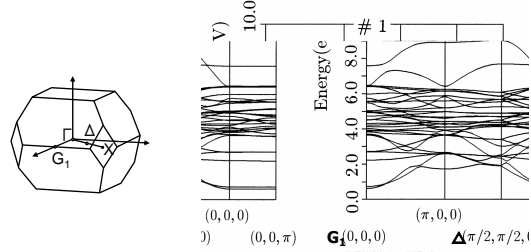} 
\caption{\label{fig4} Many-body effect including band structure \cite {KU, UK} for up-spin (or down-spin) dopant holes in underdoped LSCO above $T_{\rm c} 
$. The highest occupied band is marked by the \#1 band 
(right) and the AF Brillouin zone (left). The $\Delta$ point corresponds to $(\pi /2a, 
\pi /2a,0)$, 
while the G$_1$ point corresponds to $(\pi /a,0,0)$, at which a saddle-point singularity appears. 
} 
\end{center} 
\end{figure}

In undoped La$_{2}$CuO$_{4}$, all the energy bands in Fig. 4 are fully occupied 
by electrons so that La$_{2}$CuO$_{4}$ is an AF Mott insulator, 
consistent with experimental results. In this respect, the present effective energy band 
structure is completely different from the structure of ordinary LDA energy bands \cite{Mattheiss,Yu}. 
When Sr is doped, holes begin to occupy the top of the highest band in Fig.~\ref{fig4} 
marked by \#1 at the $\Delta$ point, which corresponds to $(\pi /2a, \pi /2a, 0)$ in the AF 
Brillouin zone. At the onset concentration of superconductivity, the Fermi level is 
located slightly below the top of the \#1 band at $\Delta$, which is  slightly higher than 
the G$_1$ point. Here, the G$_1$ point in the AF Brillouin zone lies  at $(\pi /a, 
0, 0)$ and corresponds to a saddle point of the van Hove singularity. 

\begin{figure} 
\begin{center} 
\includegraphics[width=9cm]{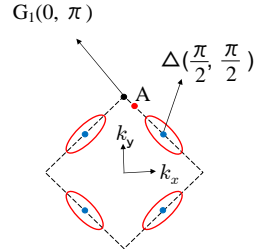} 
\caption{\label{fig5} (Color online) Fermi surface (FS) for up-spin (or down-spin) dopant holes in underdoped LSCO above $T_{\rm c}$. The FS consists of four Fermi pockets around the $\Delta$ point, $(\pi /2a, \pi /2a, 0)$, and the other three equivalent points (the nodal region) in momentum space. The figure shows the two-dimensionally projected Fermi pockets. The point A represents one of the electronic states.
} 
\end{center} 
\end{figure} 

On the basis of the calculated band structure shown in Fig. 4, Kamimura and 
Ushio  \cite{KU, UK} calculated the Fermi surface (FS) 
for the underdoped regime of LSCO. The calculated FS in the underdoped regime is composed of four Fermi pockets of extremely flat tubes. The 
projected two-dimensional (2D) picture of the four Fermi pockets around the $\Delta$ point, $(\pi /2a, \pi /2a) 
$ and the other three equivalent points in the momentum space is shown in the 
antiferromagnetic Brillouin zone in Fig. 5. The total volume of the four Fermi pockets is 
proportional to the concentration of the doped hole carriers. Thus, the feature of Fermi 
pockets constructed from the doped holes shown in Fig. 5 is consistent 
with Luttinger's theorem in the presence of AF order \cite{Luttinger}. 

In 1996 and 1997, respectively, Mason {\it et al.} \cite{Mason} and Yamada {\it et al.} \cite{Yamada} 
independently reported the magnetic coherence effects on the metallic and superconducting 
states in underdoped LSCO, determined by neutron inelastic scattering measurements.  Since then, a 
number of papers suggesting the coexistence of local AF order and superconductivity in 
cuprates as a result of neutron and NMR experiments have been published \cite{Yamada_1998, Kao, 
Christensen, Tranquada, Haydon, Mukuda}.

The Fermi surface structure in Fig. 5 is completely different from that 
of the single-component theory, in which the FS is large. Recently, Meng {\it et al.} have reported 
the existence of the Fermi pocket structure in the ARPES measurements 
of underdoped Bi$_2$Sr$_{2-x}$La$_x$CuO$_{6+\delta}$ (La-Bi2201) \cite{Meng}. Their 
results are clear experimental evidence of our Fermi pocket 
structure for underdoped LSCO predicted in 1994 \cite{KU}. 

In 1997, Anisimov {\it et al.} calculated the energy band structure 
of the ordered alloy La$_{2}$Li$_{0.5}$Cu$_{0.5}$O$_{4}$ by the LDA+U method \cite 
{Anisimov}, 
and they showed that a fairly modest reduction in the apical Cu-O bond length is 
sufficient to stabilize 
Hund's coupling spin triplet state with dopant holes in both b$_{\rm 1g}$ and a$^ 
{\ast}_{\rm 1g}$ orbitals. Their calculated result supports the K-S model. 

\section{Calculation of ARPES Spectra Based on the K-S 
Model and the Conclusion of the Absence of a Pseudogap}
Recently, considerable attention has been paid to the phenomenological idea of the 
pseudogap. When a portion of the Fermi surface in cuprates was not observed in the ARPES 
experiments, the idea of the pseudogap was proposed as a type of gap for truncating the FS 
in a single-particle spectrum \cite{Marshall, Norman_Nature}. The disconnected 
segments of the FS are called the ``Fermi arc'' \cite{Yoshida1, Norman_Nature, 
Norman_PRB2007}. Further ARPES experiments showed that such a pseudogap develops below a 
temperature denoted $T^*$, which depends on the hole concentration $x$ in the underdoped 
regime of cuprates; thus, we write $T^*(x)$ hereafter. $T^*(x)$ decreases with 
increasing hole concentration $x$ and disappears at a certain concentration $x_{\rm 
o}$ in the overdoped region \cite{Kanigel}. In this section, on the basis of the K-S 
model, we clarify the origins of the pseudogap and $T^*(x)$. 

\subsection{Calculation of the photoemission intensity and clarification 
of the origin of the observed two distinct gaps}

Below $T_{\rm c}$, the hole 
carriers in the Fermi pockets shown in Fig.5 form Cooper pairs, contributing to the formation of a 
superconducting state, and a superconducting gap appears across the Fermi level. This 
feature is consistent with Uemura {\it et al.}'s plot \cite{Uemura}. In Fig. 6(a), the 
d-wave node below $T_{\rm c}$ predicted by the K-S model \cite{Kamimura_d-Wave, 
Kamimura_Super} is schematically shown as dots, and the d-wave superconducting density of 
states is schematically shown in Fig. 6(b). Here, note 
that the AF order still coexists with a superconducting state below $T_{\rm c}$ so that we 
can use the same AF Brillouin zone, as shown in Fig. 6.

\begin{figure} 
\begin{center} 
\includegraphics[width=9cm]{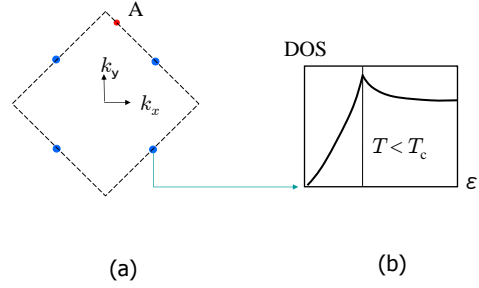}
\end{center} 
\caption{ 
\label{fig6} 
(Color online) Sketch of d-wave superconductivity in  K-S model below $T_{\rm c}$.
(a) Change of Fermi pockets in nodal region to d-wave nodes below $T_{\rm c}$. (b) d-Wave 
superconducting density of states.}  
\end{figure} 

On the other hand, in the antinodal region, the states occupied by electrons that do not 
participate in the formation of superconductivity still exist below $T_{\rm c}$. As an 
example of such states, the state A is shown in Fig.~\ref{fig6}(a), and the 
state corresponding to A above $T_{\rm c}$ is also shown in Fig.~\ref{fig5}.   
Then, real transitions of electrons from the occupied states, say, the state A, below 
the Fermi level $\varepsilon _{\rm F}$ in the \#1 energy band in 
Fig.~\ref{fig4} to a 
free-electron state above the vacuum level occur by photoexcitation both above and below 
$T_{\rm c}$ around the G$_1$ point $(\pi /a,0,0)$ and other equivalent points in momentum space. These transitions appear in the antinodal region in momentum space. 

Such a transition is shown in Fig. 7 \cite {Takahashi}. Let us 
consider the case in which an electron in the occupied state A with energy 
$\varepsilon _i$ and momentum ${\Vec{k}}_i$ in the $\#1$ energy band below $\varepsilon _{\rm F}$ is excited 
to a free-electron state with the bottom of the energy dispersion at $(\varepsilon _{\rm F} - \varepsilon _{\rm o})$ in a crystal by a photon with energy $h\nu$, where we use the suffix $i$ to emphasize the initial state of the 
transition in the crystal. Thus, the final state of the transition with energy $\varepsilon _{\rm f}$ in the crystal is expressed as 

\begin{eqnarray} 
 \varepsilon_{\rm f} =  \frac{\hbar^2}{2m} (k_{\parallel}^2 + k_{\perp}^2) + 
 \varepsilon_{\rm F} - \varepsilon_{\rm o}, \label{eq:E} 
\end{eqnarray} 
where $k_{\parallel}$ and $k_{\perp}$ are the momenta of the photoexcited electron parallel 
and perpendicular to the crystal surface, respectively. As shown in Fig. 7, the energy conservation 
for this excitation process from the initial state $|{i\rangle}$ to the final state $|{f\rangle}$ in the crystal is expressed 
as $h\nu = \varepsilon_{\rm f} - \varepsilon_{\rm i}$ for a photon of energy $h\nu$. 
                    
\begin{figure}[h] 
\begin{center} 
\includegraphics[width=9cm]{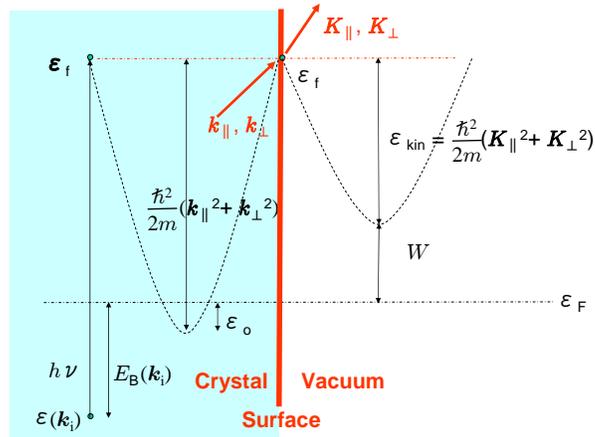}
\end{center} 
\caption{\label{fig7} 
(Color online) Energetics of photoemission process. 
} 
\end{figure} 
                       
When an electron is ejected into the vacuum level of the crystal by a photon with energy ${\hbar\nu}$, it acquires kinetic energy. Through ARPES experiments, we measure the kinetic energy of photoelectrons emitted in vacuum. We define the kinetic energy of such photoelectrons emitted in 
vacuum as $\varepsilon_{\rm kin}$, where 
 $\varepsilon_{\rm kin} =  (\hbar^2/2m) (K_{\parallel}^2 + K_{\perp}^2)$.
 
 On the other hand, the binding energy of the electron in the initial state, $E_{\rm B}$, is introduced as a new variable instead of $\varepsilon (k_{\rm i})$. $E_{\rm B}$ is defined as  
\begin{eqnarray}
  E_{\rm B} = - \varepsilon (k_{\rm i}) + \varepsilon_{\rm F}.  \label{eq:F}
\end{eqnarray}

The following equation also holds for $\varepsilon_{\rm f}$: 
\begin{eqnarray} 
 \varepsilon_{\rm f} =  \varepsilon_{\rm kin} + W + \varepsilon_{\rm F}, \label{eq:G} 
\end{eqnarray} 
where $W$ is the work function of the crystal (see Fig. 7).  

In the ARPES experiment, when a photoelectron is emitted from a crystal in vacuum 
through a surface, it is assumed that the momentum parallel to the surface is conserved: 
                          $k_{\parallel}  =  K_{\parallel}$.                          
Now, note that the $\#1$ band in Fig. 4 has been calculated by the mean field 
approximation for the fourth term in the Hamiltonian (2). The important consequence 
of this approximation is that, having taken into account the strong spin exchange 
interaction in the mean field approximation, the probability of removing an electron in the state with momentum $\Vec{k}_{\rm i}$ and energy $\varepsilon_{\rm i}$ in the $\#1$ energy band to the free-electron  state in vacuum  
can be treated in a framework similar to that for single-particle photoexcitation. 

As a result, the EDCs in the ARPES experiments corresponding to the transition from the 
occupied states in the many-body effect including energy band in Fig. 4 
to the free-electron band can be calculated using the following formula for the photoemission intensity 
$I(\Vec{k}, \omega)$:
 
\begin{eqnarray} 
               I(\Vec{k}, \omega) = \left|M_{\Vec{k}}\right|^2A(\Vec{k}, \omega)\rho_{\rm f}(\varepsilon_{\rm kin}). \label{eq:H}
\end{eqnarray} 
Here, $A(\Vec{k}, \omega)$ is the spectral function that gives the probability of removing or adding an electron at $(\Vec{k}, \omega)$, where $\omega$ is the electron energy relative to the Fermi level. It is related to the imaginary part of the one-electron Green's function; $A(\Vec{k}, \omega)  = - (1/{\pi}){\rm Im} G (\Vec{k}. \omega)$. Furthermore, $\rho_{\rm f}(\omega)$ is the density of final states and $\left|M_{\Vec{k}}\right|^2$ is the squared one-electron transition matrix element\cite {Damascelli}. It is clear from Fig. 7 that $A(\Vec{k}, \omega)$ gives the highest probability when $\hbar\omega$ is equal to $(h\nu - \varepsilon_{\rm kin} - W)$, where $h\nu$ = $\varepsilon (k_{\rm f})$ - $\varepsilon (k_{\rm i})$. 

By taking account of the lifetime effects due to the finite size of a metallic state, the deviation from the mean field approximation, and other factors, the spectral function $A(\Vec{k}, \omega)$  is given by 
\begin{eqnarray} 
               A(\Vec{k}, \omega) = (1/\pi) \frac{\delta}{[\hbar\omega - (\varepsilon_{\rm F} - \varepsilon(\Vec{k_{\rm i})})]^2 + \delta^2} ,     \label{eq:I}
\end{eqnarray} 
where $\delta$ denotes the lifetime effects, and the momentum dependence in $\varepsilon_{\rm i}$ is expressed as $\varepsilon(\Vec{k_{\rm i}})$ explicitly.

The density of final states $\rho_{\rm f}(\varepsilon_{\rm kin})$ in the EDCs is defined from 
the dispersion of the momentum of a photoexcited electron perpendicular to the crystal 
surface in the crystal, $k_{\perp}$,  as,
\begin{eqnarray} 
  \rho_{\rm f}(\varepsilon_{\rm kin}) =  \left|\frac{{\rm d}k_{\perp}}{{\rm d}
 \varepsilon_{\rm kin}}\right|. \label{eq:J} 
\end{eqnarray} 

Using eqs. (4) and (6) with the conservation of momentum of a photoexcited electron parallel to the crystal surface, $k_{\parallel}$  =  $K_{\parallel}$, the density of final states is obtained as, 

\begin{eqnarray} 
  \rho_{\rm f}(\varepsilon_{\rm kin}) = \frac{1}{2\, \sqrt{  (\hbar^2/2m) \left(\varepsilon_{\rm kin} + V  - (\hbar^2 /2m)k_{\parallel}^2\right)}}, 
  \label{eq:K}
\end{eqnarray} 
where $V$ = $W$ + $\varepsilon_{\rm o}$ is the inner potential. This result agrees with the result derived by Mizokawa \cite{Mizokawa}.

Since much of the ARPES EDC data is expressed as a function of the binding energy $E_{\rm B}$, we express eqs. (8) and (10) in terms of $E_{\rm B}$. For this purpose, we first insert eq.~(5) into eq.~(8), and simultaneously replace $\hbar\omega$ in eq.~(8) by $(h\nu - \varepsilon_{\rm kin} - W)$.  As a result, eq. (8) can be written as,
\begin{eqnarray} 
               A(\Vec{k}, \omega) = (1/\pi) 
\frac{\delta}{[(h\nu - \varepsilon_{\rm kin} - W) - E_{\rm B}]^2 + \delta^2}.    \label{eq:L}
\end{eqnarray} 

Furthermore, using the expression for the inner potential, $V$ = $W$ + $\varepsilon_{\rm o}$, and the energy conservation relation in Fig.7 given as
\begin{eqnarray}
   h\nu =  \varepsilon_{\rm kin} + W + E_{\rm B}, \label{eq:M}
\end{eqnarray}
eq. (10) can be expressed as
\begin{eqnarray} 
  \rho_{\rm f}(\varepsilon_{\rm kin}) = \frac{1}{2\, \sqrt{  (\hbar^2/2m) \left(h\nu - E_{\rm B} + \varepsilon_{\rm o} - (\hbar^2 /2m)k_{i\parallel}^2\right)}}, \label{eq:M}
\end{eqnarray} 
where $k_{i\parallel}$ is the component of ${\Vec{k}}_i$ parallel to the crystal surface.

  Using eqs. (7), (12), and (13), we have calculated the photoemission intensity $I(\Vec{k}, \omega)$ as a function of $E_{\rm B}(\Vec{k_{\rm i}})$. In performing the numerical calculations, we have considered that the photon energy ($h\nu$) range in synchrotron radiation experiments is 10 to 100 eV and the kinetic energy range of the photoelectron is also 10 to 100 eV \cite {Gweon}. Since the width of the energy dispersion of the $\#1$ energy band in Fig. 4(a) is about 1 eV, we notice that the $E_{\rm B}$ range is up to 1 eV. For $\delta$, whose inverse gives a measure of the lifetime broadening in the $\#1$ band, we assume 100 meV on the basis of the discussion in the subsequent section.

In this context, we choose 15 eV for $h\nu$, 10 eV for $\varepsilon_{\rm kin}$, 3.5 eV for W \cite {Kinoda}, and 4.5 eV for the inner potential $V$ = $\varepsilon_{\rm o}$ + $W$ in the present numerical calculations. As regards 
$\hbar^2k_{i\parallel}^2/2m$, we choose the center of the antinodal region, 
i.e., the $G_1$ point or the edge of the AF Brillouin zone in Fig. 4, for the $i$th point, because the antinodal region is narrow around the $G_1$ point, so that $\hbar^2k_{i\parallel}^2/2m$ does not change much upon varying the $i$th point. By adopting the empty lattice test for the free-electron energy bands, we estimate $\hbar^2k_{i\parallel}^2/2m$ to be 3 eV for $i$ = $G_1$. 

The calculated $I(\Vec{k}, \omega)$ with the values of the above parameters is shown as a function of $E_B$ in Fig. 8(a). Since $\varepsilon_{\rm kin}$ is very large, a divergent point in the density of final states $\rho_{\rm f}(\omega)$  appears at a large $E_{\rm B}$. Thus, the photoemission intensity $I(\Vec{k}, \omega)$ shows a feature of a broad hump, reflecting a peak in the spectral function $A(\Vec{k}, \omega)$ given by eq. (11), as seen in Fig. 8(a). This trend is consistent with the experimental results of the ARPES spectra of underdoped Bi2212 samples below $T_{\rm c}$ in the antinodal region by Tanaka {\it et al.} \cite{Tanaka}, although the shape of the broad hump is slightly different.

\begin{figure}[h] 
\begin{center} 
\includegraphics[width=9cm]{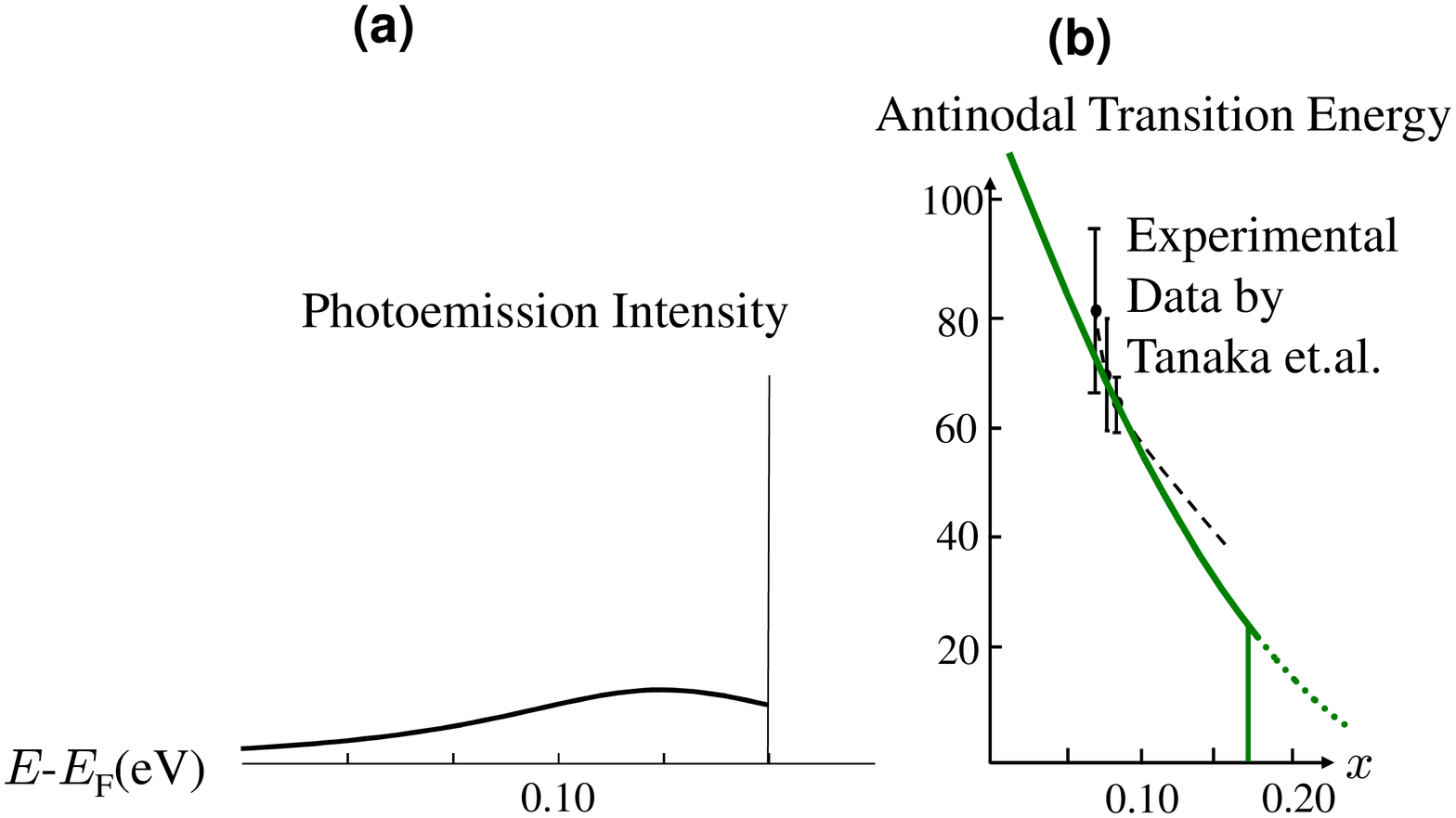}
\end{center} 
\caption{\label{fig8} 
(Color online) Calculated ARPES spectra of LSCO and their comparison with experimental results of Bi2212. 
(a) Calculated photoemission intensity as a function of binding energy $E_B (= E - E_{\rm F})$. (b) Calculated energy difference $|\varepsilon ({\rm G}_1)-\varepsilon _{\rm F} 
(x)|$ (antinodal transition energy) as a function of hole concentration $x$ at $T = 0$K. Experimental results of 
Tanaka {\it et al.} \cite{Tanaka} are shown by dots. 
} 
\end{figure} 

From the ARPES spectra in the nodal region shown in Fig. 6(b), which was predicted from the d-wave superconductivity due to the K-S model, \cite{Kamimura_d-Wave} and those in the antinodal region shown in Fig. 8(a), we can conclude that the features of ARPES spectra below $T_{\rm c}$ are 
theoretically as follows: ARPES spectra consist of a coherent peak due to the superconducting density of states that appears in the nodal region around the $\Delta$ point and a broad hump 
that appears in the antinodal region, which corresponds to 
the real transition of electrons from the occupied states below 
the Fermi level $\varepsilon _{\rm F}$ to a free-electron state above the vacuum level.   These theoretical results of ARPES EDCs are similar to the experimental ones
 reported by Tanaka {\it et al.} for Bi2212 \cite{Tanaka}, where the experimental 
results revealed two distinct energy gaps in the nodal and antinodal regions exhibiting different doping dependences. Thus, we designate the broad hump in the antinodal region as an ``antinodal transition'', 
where $\varepsilon _{\rm F}$  varies with the hole concentration $x$, 
 so that we write $\varepsilon _{\rm F}(x)$ hereafter. 

Furthermore, Tanaka {\it et al.} reported the doping dependence of the position of the hump, which is determined from the second derivative of the spectra in the antinodal region around the $G_1$ point for the three underdoped samples in Bi2212. We compare this experimental result of Bi2212 with the calculated doping dependence of the binding energies $E_B$ at the $G_1$ point for LSCO, which correspond to the energy difference 
$|\varepsilon({\rm G}_1)-\varepsilon_{\rm F}(x)|$. We call this energy difference the ``antinodal transition energy''. 

Since the shape of the density of states (DOS) for the highest 
conduction band does not depend on the type of cuprate material, we can compare in Fig.~\ref{fig8}(b) the calculated doping dependence of the antinodal transition energy for LSCO (solid 
lines) with the experimental results of the antinodal gap of Bi2212 in ref.16, which are 
shown as dots in the figure. As seen in Fig.~\ref{fig8}(b), the agreement between the 
theory and the experiment is remarkably good. From this quantitative agreement, we can 
conclude that among the observed two gaps below $T_{\rm c}$, the gap associated with the 
antinodal regime corresponds to the real transitions of electrons from the occupied 
states below the Fermi level to a free-electron state above the vacuum level, while the 
other gap associated with the near-nodal regime corresponds to the superconducting gap 
created on Fermi pockets. From the excellent agreement between the present theoretical results and the 
experimental results of Tanaka {\it et al.}, we can conclude that the real 
transitions of electrons from occupied states below the Fermi level to a 
free-electron state above the vacuum level by photoexcitation appear in the antinodal region in underdoped 
cuprates so that the introduction of a pseudogap is not necessary. 

Recently, Yang {\it et al.} \cite{Yang} have suggested from their ARPES experiments on 
Bi2212 that the opening of a symmetric gap related to superconductivity occurs only in 
the antinodal region and that the pseudogap reflects the formation of preformed pairs, 
in contrast to the ARPES experimental results reported by Tanaka {\it et al} \cite{Tanaka}. 
In the present theory, we have clearly shown in Figs.~\ref{fig6} and \ref{fig8} that, 
in the ARPES experiments, a peak related to superconductivity appears only in the nodal 
region and that the spectra in the antinodal region correspond to photoexcitations from 
occupied states below $\varepsilon_{\rm F}$ to a free-electron state above the 
vacuum level. If the antinodal region in ref. 44 is the region around the ${\rm G}_1$ point 
in the present paper, their suggestion is in disagreement with our theoretical results. Finally, 
we should remark that any proposed theory must explain both the doping and temperature 
dependences of ARPES spectra in the underdoped regime consistently. From this standpoint, 
we will investigate the temperature dependence of the calculated broad hump in ARPES EDCs theoretically in the 
next subsection.

\subsection{Physical meaning of $T^* (x)$ and the temperature 
dependence of ARPES spectra}

To calculate the temperature dependence of the antinodal transition 
energy, first we would like to clarify the physical meaning of $T^* (x)$. When a 
hole concentration $x$ is fixed at a certain value in the underdoped region and 
the temperature increases beyond $T_{\rm c}$, in the normal phase, the local AF order 
constructed by superexchange interaction in a CuO$_2$ plane is destroyed by thermal 
agitation, and thus a phase showing the coexistence of a metallic state with the Fermi 
pockets and the local AF order diminishes gradually. As a result, an electronic 
phase consisting of a large FS without the AF order is mixed with a phase of the K-S 
model. Finally, at a certain temperature, a uniform phase consisting of the electronic 
phase consisting of a large FS without the AF order will appear in the underdoped regime. 
This temperature is defined as $T^* (x)$. Thus, the phase of the Fermi pockets coexisting 
with the local AF order in the K-S model holds only below $T^* (x)$. We designate the 
phase of the Fermi pockets in the K-S model as the ``small FS'' phase and the electronic 
phase consisting of a large FS without the AF order as the ``large FS'' 
phase. Hereafter, the former and latter are abbreviated as the SF and LF phases, 
respectively. In this context, one may consider that a phase below $T^* (x)$ is a mixed 
phase of the SF and LF phases in the underdoped regime; thus, $T^* (x)$ 
represents a crossover from the mixed phase to the LF  phase. 

To calculate $T^* (x)$ on the basis of the K-S model, one must take account 
of the effect of thermal agitation in the system of Cu localized spins in the 
AF order (the first story in Fig. 2). However, such calculation 
is possible only for a finite system, as Hamada and coworkers have shown 
\cite{Hamada, KH}. In this context, we calculate $T^* (x)$ approximately, 
neglecting the effect of thermal agitation in the system of Cu localized 
spins.

For this purpose, 
let us introduce a quantity that defines the difference between the free energies of pure LF 
and SF phases: 
\begin{eqnarray} 
\Delta F(T, x)&\equiv &F_{\rm LF}(T, x)-F_{\rm SF}(T, x), 
\label{eq1} 
\end{eqnarray} 
where $F_{\rm LF}(T, x)$ and $F_{\rm SF}(T, x)$ are the free energies of the LF and SF  
phases, respectively. Here the free energy $F(T, x)$  is defined as 
\begin{eqnarray} 
F(T, x)&=&E (T, x)-TS(T, x), 
\label{eq2} 
\end{eqnarray} 
where $E(T, x)$ and $S(T, x)$ are the internal energy and entropy of each phase, 
respectively. These quantities are calculated from 
\begin{eqnarray} 
E(T, x) &=& \int ^{\infty}_{-\infty}\varepsilon \rho (\varepsilon ) f(\varepsilon , \mu 
(x)) {\rm d}\varepsilon , 
\label{eq3a} 
\end{eqnarray} 
and 
\begin{eqnarray} 
S(T,x) &=& - k_{\rm B} \int^{\infty}_{-\infty} \Bigl[ f(\varepsilon,\mu(x)) \ln f 
(\varepsilon,\mu(x)) \nonumber \\ 
      & & {}+ \bigl\{ 1-f(\varepsilon,\mu(x)) \bigr\} \ln \bigl\{ 
1-f(\varepsilon,\mu(x)) \bigr\} \Bigl]~ 
\nonumber \\ 
       & & ~~~~~~~~~~~~~~~~~~~~~~~~~~~~~~~~~~~~~~ \rho(\varepsilon)~ d\varepsilon \label 
{eq:S} \,, 
\label{eq3b} 
\end{eqnarray} 
where $\mu (x)$ is the chemical potential of each phase, $\rho $($\varepsilon $) is the 
 DOS for each phase, and $f(\varepsilon , \mu(x))$ is the Fermi distribution 
function at energy $\varepsilon$ and chemical potential $\mu (x)$. Then, $\tilde{T}^* (x) 
$ is defined by 
\begin{eqnarray} 
\Delta F(\tilde{T}^* (x), x) = 0. 
\label{eq4} 
\end{eqnarray} 

\begin{figure}[h] 
\begin{center} 
\includegraphics[width=9cm]{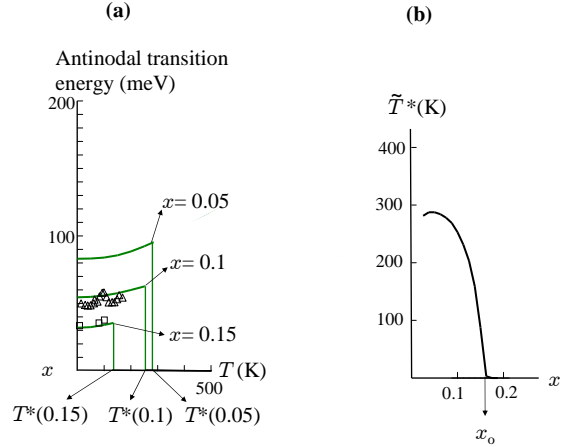} 
\end{center} 
\caption{\label{fig9} 
(Color online) (a) Calculated temperature dependence of 
antinodal transition energy for $x$ = 0.05, 0.1, and 0.15. Experimental data obtained by Norman 
{\it et al.} \cite{Norman_Phys_Rev} (triangles) and Lee {\it et al.} \cite{Lee} (squares) are 
shown. (b)  Calculated result of $\tilde{T}^* (x)$ as a function of the hole concentration $x$}
\end{figure} 

Kamimura {\it et al.} calculated the electronic entropies for the SF and LF phases 
of LSCO \cite{Kamimura_Entropy}. According to their results, the difference in electronic 
entropy between the SF and LF phases increases with increasing hole concentration 
$x$ in the underdoped regime. Using this result, we have calculated   
$\tilde{T}^* (x)$ from eq.~(\ref{eq4}) as a function of $x$, instead of $T^* (x)$. In doing so, we have 
introduced two parameters, $\tilde{T}^*(x=0.05)$ and $x_{\rm o}$, where $x_{\rm o}$ 
is the critical concentration that satisfies $\tilde{T}^*(x_{\rm o})$=0. 
Here, 
$\tilde{T}^*(x=0.05)$ represents a quantity related to the energy difference between the 
phase of the doped AF insulator and the LF phase at the onset concentration of the metal- 
insulator transition ($x=0.05$). $\tilde{T}^*(x=0.05)$ is chosen to be 300K. On the other hand, the physical meaning of $\tilde{T}^*(x_{\rm o})=0$ can be 
explained as follows: When the hole concentration exceeds the optimum doping level 
($x$ = 0.15) for LSCO and enters a slightly overdoped region, the local AF order via 
superexchange interaction in a CuO$_2$ plane is destroyed by an excess of hole 
carriers. Thus, the K-S model does not hold at a certain concentration $x_{\rm o}$ in the 
overdoped region, and hence the small FS in the K-S model changes to a large FS. Thus, 
$\tilde{T}^* (x)$ vanishes at $x_{\rm o}$. From the analysis of various experimental 
results, we choose $x_{\rm o}$ = 0.17 for LSCO \cite{Nakano, Cooper}. The calculated 
result of $\tilde{T}^*(x)$ with $\tilde{T}^*(x=0.05)$ = 300K and 
$x_{\rm o}$ = 0.17 is shown as a function of $x$ in Fig.~\ref{fig9}{\bf b}.

From the present result, we can say that the area below $\tilde{T}^* (x)$ in the 
underdoped regime represents the region in which the normal (metallic) phase above $T_{\rm 
c}$ and the superconducting phase below $T_{\rm c}$ coexist with the local AF order. In a 
real system, a region of a mixed phase consisting of the SF and LF phases appears between 
$\tilde{T}^* (x)$ and 
$T^* (x)$ owing to the dynamical interaction of the fourth term in the effective 
Hamiltonian~(\ref{eq:2}). Thus, $T^* (x)$ always appears 
above $\tilde{T}^* (x)$ \cite {Miyakawa}.

 Under this circumstance, it is clear that the antinodal transition energy defined by   
$|\varepsilon({\rm G}_1)-\varepsilon_{\rm F}(x) |$  appears at temperatures below $T^*(x) 
$ and vanishes at $T^*(x)$. By using $\tilde{T}^*(x)$ instead of $T^*(x)$, we calculate 
the temperature dependence of the antinodal transition energy using eqs. (\ref 
{eq1})-(\ref{eq3b}). The calculated results for three 
concentrations, i.e., $x$ = 0.05, 0.10, and 0.15, of LSCO in the underdoped-to-optimallydoped 
region are shown in Fig.~\ref{fig9}(a) as functions of temperature, where $\tilde{T}^*(0.05)= 300$K is used.  As seen in the figure, the antinodal 
transition energy increases slightly with temperature up to $\tilde{T}^*(x)$ and vanishes 
suddenly at $\tilde{T}^*(x)$. These calculated results are compared with the experimental 
results of the underdoped sample of Bi2212 in refs. 47 and 48, which are indicated by triangles 
and squares, respectively, in Fig.~\ref{fig9}(a). As seen in the figure, the 
agreement between the theory and the experiment is remarkably good, although the present 
theoretical treatment is not rigorous in the sense that, in the calculation of $F_{\rm 
SF}(T, x)$, the dynamical interplay of a metallic state (the second story in 
Fig.~\ref 
{fig2}) and Cu localized spins in the AF order (the first story) is not taken into 
account. 

\section{Spatially Inhomogeneous Distribution of Fermi Pocket States and Large Fermi Surface States due to the Finite Size Effects} 
\subsection{Finite size effects of metallic state on Fermi surface} 
According to the results of neutron inelastic scattering experiments by 
Mason {\it et al.} \cite{Mason} and Yamada {\it et al.} \cite{Yamada}, the AF spin-correlation length $\lambda_{s} 
$ in the underdoped region of LSCO is finite. In the underdoped regime of LSCO, it 
increases as the Sr concentration increases from $x = 0.05$ in LSCO, 
the onset of superconductivity, and reaches about 50\AA ~or more at the 
optimum doping level ($x = 0.15$). In this subsection, we discuss the effects of 
the finite size of the AF spin-correlation length on the structure of the Fermi pockets shown 
in Fig. 5. According to the K-S model in Fig.~\ref{fig2}, in the spin-correlated 
region a doped hole in the underdoped regime of LSCO can itinerate coherently by taking 
the a$^*_{\rm 1g}$ and b$_{\rm 1g}$ orbitals alternately in the presence of the local AF 
order without destroying the AF order.

In the case of a finite spin-correlated region, one may think that there are 
frustrated spins at the boundary between the spin-correlated region of the AF order and 
the region of the ``resonating valence bond'' (RVB) state proposed by Anderson \cite 
{Anderson} without hole carriers. Here, the frustrated spins mean that the localized spins at the boundary are 
not in the AF order, but directed parallel to each other. Suppose that one of the frustrated 
spins in a parallel direction at the boundary  changes  its direction from parallel 
to antiparallel by the fluctuation effect in the 2D Heisenberg AF spin system  during the 
time of $\tau_{s}$ defined by $\tau_{s} \equiv \hbar /J$, 
where $J$ is the superexchange interaction ($\sim$0.1 eV). 
At the time of $\tau_{s}$, on the other hand, hole carriers at the Fermi level can 
move with the Fermi velocity inside the spin-correlated region of the AF order. The 
traveling time of a doped hole at the Fermi level 
over an area of the spin-correlation length 
is given by $\tau_{\rm F} \equiv \lambda_{s}/v_{\rm F}$, 
where $v_{\rm F}$ is the Fermi velocity of a doped hole at the Fermi level. 
In the case of underdoped LSCO, $\tau_{s}$ is 6 $\times$ $10^{-15}$ s. 
Since $v_{\rm F}$ is estimated to be 2.4 $\times$ $10^{4}$ m/s from the dispersion of 
the \#1 band in Fig.~\ref{fig4},  $\tau_{\rm F}$ is 2 $\times$ $10^{-13}$ s 
for the underdoped region of $x=0.10$ to $x=0.15$ in LSCO, where for the spin-correlation 
length $\lambda_{s}$ at $x = 0.15$, we have chosen 50~\AA. 
Thus, $\tau_{\rm F}$ 
becomes much longer than $\tau_{s}$. As a result, the frustrated spins on the boundary 
change their directions from parallel to antiparallel before a hole carrier in the spin-correlated region of the AF order reaches the boundary. Thus, a metallic state for a doped 
hole becomes much wider than the observed spin-correlation length by the passing of a doped hole through 
the boundary without spin scattering. In this way, a metallic state is surrounded by RVB states and the spatial 
distribution of the metallic states 
is inhomogeneous. 

Finally, we explain why we have chosen 100 meV for $\delta$ in calculating the photoemission intensity shown in Fig.~\ref{fig8}(a). For example, the initial state of photoexcitation in ARPES near the $G_1$ point is either a component of Hund's coupling triplet $^3$B$_{\rm 1g}$ or Zhang-Rice singlet $^1$A$_{\rm 1g}$ shown in Fig.~\ref{fig3} in the \#1 band \cite{UK}. Thus, if 
the local AF order between neighboring Cu sites in Fig.~\ref{fig3} is 
destroyed, the calculated result in Fig.~\ref{fig8}(a) may not be valid. This is
 the reason why $\delta$ is the same order of magnitude as the inverse of 
$\tau_{s}$, that is,  the superexchange interaction $J$ ($\sim$0.1 eV).

\subsection{Origin of the coexistence of a local AF order and a metallic 
state: The kinetic-energy-driven mechanism} 

 Concerning the finite system of cuprates, Hamada {\it et al.}  \cite{Hamada} and Kamimura 
and Hamada \cite{KH} attempted to determine the ground state of the effective Hamiltonian~ 
(\ref{eq:2}) for the K-S model by carrying out the exact diagonalization of the 
Hamiltonian ~(\ref{eq:2}) using the Lanczos method for a 2D square 
lattice system with 16 (4 $\times$ 4) localized spins with one and two 
doped holes, respectively. As a result they clarified 
that, in the presence of hole carriers, the localized spins in a spin-correlated region 
tend to form an AF order rather than a random spin-singlet state, and 
thus hole carriers can lower the kinetic energy by itinerating in the lattice 
of the AF order (the first story in Fig.2). This is the mechanism 
leading to the coexistence of a 
metallic state and a local AF order in the K-S model. 

Generally, a hole-carrier in the spin-correlated region of the AF order can 
propagate through the boundary of the spin-correlated region with the above-mentioned mechanism of the K-S model; hence, the region of a metallic state coexisting with the AF order 
becomes much wider than the observed spin-correlated region.  In fact, Kamimura {\it et al.}  estimated the length of the metallic region at the optimum doping level of LSCO to be 
about 300~\AA~ from the $T_{\rm c}$ at the optimum doping level \cite{Kamimura_Super}.   
Recently, an idea similar to ours with regard to the decrease in the kinetic energy has 
been proposed by Wrobel and coworkers, who have shown that the decrease in the kinetic 
energy is the driving mechanism that induces superconductivity \cite{Wrobel, 
Wrobel_Furde}.

\section{Remark on a Phase Diagram for Underdoped Cuprates} 

\begin{figure}[h] 
\begin{center} 
\includegraphics[width=9cm]{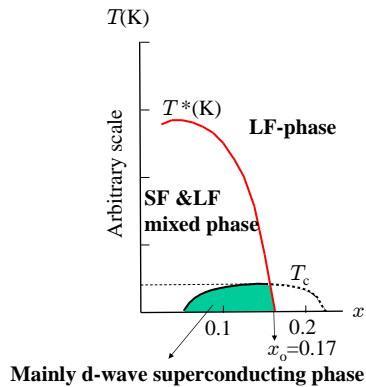} 
\end{center} 
\caption{\label{fig10} 
(Color online) New explanation for phase diagram of LSCO. Since $T^*(x)$, which is higher than 
$\tilde{T}^*(x)$, has not been calculated, its numerical values are not 
shown on the vertical axis.} 
\end{figure}

From the calculated results shown in Fig.~\ref{fig9}(b), we would like to 
comment on the $T$ vs $x$ phase diagram for cuprates shown in Fig.~\ref{fig10}, for which 
it has been said that a pseudogap state exists below the temperature $T^*(x)$. According 
to our calculations in previous sections, 
the SF phase constructed from Fermi pockets appears in the presence of the 
local AF order below 
$T^*(x)$ in the underdoped region. However, when the temperature increases at a fixed 
concentration in the underdoped regime, the AF order is destroyed gradually with 
increasing temperature, and thus the K-S model does not hold slightly below $T^*(x)$. 
On the other hand, when the hole concentration increases at a fixed temperature, the AF 
order is destroyed by overdoped holes. Thus, the K-S model does not hold at a certain hole 
concentration. The thermal effect and excess hole density effect cause a mixing of the SF and 
LF phases, as explained in \S 4. 

In this context, we would like to point out that the area below  $T^*(x)$ and above 
$T_{\rm c}$ in Fig.~\ref{fig10} is not a pure phase but an inhomogeneous distribution of 
Fermi pocket states and  large FS states. Thus, we can say that, when 
the temperature increases from $T_{\rm c}$ at a fixed hole concentration in the underdoped 
region, the population showing Fermi pockets decreases while that showing large 
FS increases with increasing temperature. Finally, when the temperature 
exceeds $T^*(x)$, a uniform LF phase appears as a metallic state. Thus, we may say that 
$T^*(x)$ is a crossover line from the inhomogeneous mixed phase of Fermi pockets and 
large FS to the LF phase. 

This result can explain the strange temperature evolution of a Fermi arc observed by 
Norman {\it et al.} \cite{Norman_Nature} and Kanigel {\it et al} \cite{Kanigel}. Furthermore, 
in the 
superconducting phase below $T_{\rm c}$ and $T^*(x)$ in Fig.~\ref{fig10}, an s-wave 
component of superconductivity originating from the LF phase may be mixed with the d-wave 
superconductivity. Such a mixing effect was experimentally reported by M\"{u}ller \cite 
{Mueller}. In this context, we conclude that $T^*(x)$ represents a crossover from the SF phase to the LF phase rather than a phase boundary between the pseudogap phase and a metal \cite{McElroy}. 

Furthermore we can predict that the spin susceptibility will show 2D-like AF 
features mainly below $T^*(x)$ and Pauli-like temperature-dependent behavior above 
$T^*(x)$. We find that this prediction is also consistent with the experimental results 
for LSCO \cite{Nakano,Cooper}. In this context, it should be emphasized that the K-S model 
is shown to explain successfully not only the ARPES experimental results \cite{Meng, 
Tanaka, Norman_Phys_Rev, Lee} but also a number of other experimental results such as NMR 
results showing the coexistence of a superconducting state and AF order 
\cite{Mukuda}, polarized X-ray absorption spectra \cite{Chen, Pellegrin}, site-specific 
X-ray absorption spectroscopy \cite{Merz_Nucker}, anomalous electronic entropy \cite 
{Loram_Entropy, Kamimura_Entropy}, and d-wave superconductivity \cite{Wollman_d-Wave, Tsuei}. Theoretically, the K-S model is also supported by {LDA + U} band calculations \cite 
{Anisimov}, as already mentioned in \S 2.5. 

\section{Conclusions and Concluding Remarks}

In this study we have shown on the basis of the K-S model how the interplay of Mott 
physics and JT physics plays an important role in determining the 
superconducting state as well as the metallic state of underdoped cuprates. In connection with the interplay of JT physics and Mott physics, the following 
important results have been obtained in this study: It has been clarified on the basis of 
the K-S model that the concept of the pseudogap discussed theoretically\cite{Wrobel,Schmalien, 
Yang_Rice_Zhang} and reported by ARPES, STM, and tunneling experiments below $T^*(x)$ in 
underdoped cuprates \cite{Norman_PRB2007, Kanigel, Renner} can be explained by the 
occurrence of Fermi pockets in the underdoped region without the pseudogap hypothesis. 
We have shown that the appearance of the broad hump observed in the antinodal region in 
ARPES can be explained by the real transitions of photoexcited electrons from the occupied 
states in the highest conduction band in the antinodal region to a free-electron state 
above the vacuum level. Furthermore, we have shown that the physical meaning of $T^*$ 
represents a crossover line from an inhomogeneous phase consisting of Fermi pockets 
and large FS in the normal state to a phase consisting of a large FS.

Finally, several remarks are made on the small FS and shadow bands in the 
underdoped regime of cuprates. In 1996, Wen and Lee developed a slave-boson theory for the 
t-J model at finite doping level, and showed that Fermi pockets at low doping continuously 
evolve into a large FS at high doping concentrations \cite{Xiao-Gang_Wen}. 
Although their theoretical model is different from the K-S model, it is interesting to 
find that they obtained a result similar to the result predicted using the K-S model in 1994 with 
regard to the change from a small FS to a large FS with increasing hole concentration. 
Recently, a proposal was made to reconcile the experimental result of the coexistence of 
antiferromagnetism and superconductivity \cite{Kaul}. Furthermore, in relation to the small 
FS, the idea of a shadow FS was proposed as a replica of the main 
FS transferred using $Q = (\pi/a, \pi/a)$ by Kampf and Schrieffer theoretically 
\cite{Kampf} and by Aebi {\it et al.} experimentally \cite{Aebi}. Investigating the validity of the idea 
of the shadow FS experimentally, the observation of shadow bands 
in ARPES spectra has been reported \cite{Vilk, Saini, Nakayama}. Responding to the 
problems of the shadow FS and shadow bands from the standpoint of the K-S model, it should be 
emphasized that Fermi pockets in the metallic state calculated from the K-S model 
have been derived as a result of the interplay of JT physics and Mott physics; thus, the origin of Fermi pockets is different from 
that in a single-component theory. 
Therefore, the Fermi pockets shown in Fig.~\ref{fig5} are neither the shadow FS nor related to the shadow bands. 

\begin{acknowledgments} 
We would like to thank Dr. Wei-Shen Lee, and Profs. Atsushi Fujimori, Nobuaki 
Miyakawa, Naurang Saini, Tomohiko Saitoh, Hideaki Sakata, and Xingjiang Zhou for valuable discussions 
on experimental results. We also thank Prof. Takashi Mizokawa for a helpful discussion on ARPES EDC experiments. Finally, we thank Dr. Jaw-Shen Tsai for valuable comments on the present work. 
This work was supported by the Quantum Bio-Informatics Center of Tokyo University of Science. 
\end{acknowledgments} 


\end{document}